\documentclass[11pt]{article}

\usepackage[margin=1in]{geometry}
\usepackage{microtype}
\usepackage{hyperref}
\usepackage{url}
\usepackage{booktabs}
\usepackage{enumitem}

\hypersetup{
  colorlinks=true,
  linkcolor=blue,
  urlcolor=blue,
  citecolor=blue
}

\title{Mind the Boundary: Stabilizing Gemini Enterprise A2A via a Cloud Run Hub Across Projects and Accounts}
\author{Takao Morita (Takao Morita)\\Independent Researcher}
\date{}

\begin{document}
\maketitle

\begin{abstract}
Enterprise conversational UIs increasingly need to orchestrate heterogeneous backend agents and tools across project and account boundaries in a secure and reproducible way. Starting from Gemini Enterprise’s Agent-to-Agent (A2A) invocation, we implement an A2A Hub (orchestrator) on Cloud Run that routes queries to: (i) a public A2A agent deployed in a different project, (ii) an IAM-protected Cloud Run A2A agent in a different account, (iii) a RAG path combining Discovery Engine / Vertex AI Search with direct retrieval of source text from Google Cloud Storage (GCS), and (iv) a general QA path via Vertex AI. We show that practical interoperability is governed not only by protocol compliance but also by Gemini Enterprise UI constraints and boundary-dependent authentication. Real UI requests arrive in params.message.parts[].text and include acceptedOutputModes=[], so mixing structured data into JSON-RPC responses can trigger UI errors. To address this, we enforce a text-only compatibility mode on the JSON-RPC endpoint while separating structured outputs and debugging signals into a REST tool API. On a four-query benchmark spanning expense policy, PM assistance, general knowledge, and incident-response deadline extraction, we confirm deterministic routing and stable UI responses; for the RAG path, granting storage.objects.get enables evidence-backed extraction of the “within 15 minutes” deadline. All experiments are reproducible using the repository snapshot tagged a2a-hub-gemini-ui-stable-paper.
\end{abstract}

\section{Introduction}
Enterprise adoption of generative AI is shifting from single-LLM question answering toward multi-agent architectures that connect specialized agents and business systems behind a conversational user interface (UI). In such settings, a user issues natural-language instructions, while the backend coordinates distinct capabilities such as expense handling, project management support, internal document search, and general knowledge assistance.

Google Gemini Enterprise supports custom agent integration using the Agent-to-Agent (A2A) protocol, which combines agent discovery (via an agent card) and message exchange over JSON-RPC. However, in practice, a configuration that is “allowed by the specification” does not necessarily behave stably across UI constraints and cloud boundary conditions.

In this study, we start from Gemini Enterprise UI and validate cross-boundary invocation of agents hosted on Cloud Run through an intermediate A2A Hub (orchestrator). We evaluate three boundary types: same project, different project (same account), and different account. We focus on practical issues that are under-specified in official documents, including UI compatibility, authentication design across boundaries, and permission design for Retrieval-Augmented Generation (RAG).

Our contributions are: (i) we document a concrete interoperability gap between Gemini Enterprise UI and A2A implementations based on primary request/response observations; (ii) we organize authentication design across project and account boundaries for agent invocation; and (iii) we demonstrate that a Hub-based approach improves UI stability and error containment in cross-boundary agent orchestration.

\smallskip
\noindent\textbf{Code availability.} The source code used in this study is publicly available at:
\url{https://github.com/taotaotao3/a2a-agent-hub}

\section{Background and Related Work}
Prior work on tool-augmented and agent-based LLM systems has explored
modular reasoning and orchestration approaches \cite{react, mrkl}.
\subsection{A2A protocol and agent cards}
The A2A protocol specifies a loosely coupled mechanism for agent-to-agent communication, defining a discovery interface via an agent card and a JSON-RPC message exchange interface. The agent card is typically hosted at \texttt{/.well-known/agent-card.json} and declares supported input/output modes and skills. Gemini Enterprise uses this interface to allow UI-driven invocation of externally hosted agents. Notably, the A2A specification does not prescribe authentication mechanisms or UI rendering constraints; these boundary conditions are left to deployment and client implementations.

\subsection{Multi-agent orchestration}
Prior work such as MRKL and ReAct shows how LLM-driven systems can use external tools or modules in a task-adaptive manner. In contrast, our Hub emphasizes deterministic routing over LLM-based deliberation to maximize reproducibility and debugging clarity in an enterprise setting.

\subsection{RAG in enterprise data}
RAG augments a generation model by retrieving relevant context from internal documents or knowledge bases. On Google Cloud, Discovery Engine / Vertex AI Search offers managed retrieval services. In practice, retrieval snippets and direct access to the original documents (e.g., in GCS) can require distinct IAM configurations, which becomes a key operational factor for evidence-grounded answers.

\section{System Design}
This section presents the A2A Hub architecture for achieving cross-boundary agent interoperability starting from the Gemini Enterprise UI.

\subsection{Overall architecture}
The system comprises (1) the Gemini Enterprise UI, (2) an A2A Hub on Cloud Run, and (3) multiple downstream agents and tool paths. The UI sends A2A JSON-RPC requests to the Hub; the Hub normalizes input, performs routing, and forwards requests to the appropriate downstream path. The Hub intentionally minimizes “model” responsibilities and instead focuses on (i) input normalization, (ii) deterministic routing, and (iii) absorption of boundary-dependent differences.

We implement four logical routes: (i) an expense-related public A2A agent deployed in a different project, (ii) a project-management (PM) support A2A agent deployed in a different account and protected by Cloud Run IAM, (iii) a document QA (DocQA) route combining Discovery Engine / Vertex AI Search with optional direct retrieval of source text from GCS, and (iv) a general QA route via Vertex AI. All components are deployed on Cloud Run, with HTTP used for inter-service communication.

\subsection{UI compatibility as a first principle}
We observed that Gemini Enterprise UI sends A2A requests where user input is stored in \texttt{params.message.parts[].text}, and that \texttt{acceptedOutputModes} is an empty array (i.e., \texttt{acceptedOutputModes=[]}). Under this condition, responses that mix structured JSON with text may be acceptable to the A2A specification, yet still fail UI interpretation and cause user-visible errors.

Therefore, we adopt \emph{text-only UI compatibility} as a design principle. For the JSON-RPC endpoint (\texttt{POST /}), the Hub always returns a single text part, and it avoids propagating HTTP 500 to the UI by summarizing errors as normal text responses. In parallel, structured data and rich debugging signals are separated into a REST tool API (\texttt{POST /tools/query}). This separation preserves UI stability while enabling robust inspection during development and evaluation.

\subsection{Input normalization and deterministic routing}
To tolerate client differences and potential future changes, the Hub accepts both \texttt{params.text} and \texttt{params.message.parts[].text} and normalizes inputs into a canonical text query. Routing is performed using deterministic rules (keyword/regex-based), not LLM inference. For example, queries containing Japanese terms for expense reimbursement are routed to the expense agent, while queries containing PM/WBS terms are routed to the PM agent. This design prioritizes reproducibility and fault localization.

\subsection{Absorbing boundary-dependent authentication}
While A2A does not specify authentication, Cloud Run and IAM boundaries dominate real deployments. We organize authentication by boundary type: same-project resources rely on Application Default Credentials (ADC) from the Hub’s Cloud Run service account; cross-project public agents are invoked via unauthenticated HTTPS; and cross-account agents require explicit OIDC ID tokens (with correct audience) and Cloud Run Invoker permissions granted to the Hub’s service account. The Hub hides these differences behind a uniform A2A interface to the upstream UI.

\section{Implementation}
This section describes the implementation of the Hub and its downstream integration. The system is implemented in Python as an asynchronous web application deployed on Cloud Run.

\subsection{Hub structure and endpoints}
The Hub is implemented as an ASGI application (Starlette) and served by Uvicorn. A single codebase provides both the A2A JSON-RPC endpoint and the REST tool API. The main endpoints are:
\begin{itemize}
  \item \texttt{POST /}: A2A JSON-RPC (\texttt{message/send})
  \item \texttt{POST /tools/query}: REST tool API for debugging/inspection
  \item \texttt{GET /.well-known/agent-card.json}: agent card
  \item \texttt{GET /openapi.yaml}: REST API definition
  \item \texttt{GET /health}, \texttt{/routes}, \texttt{/debug-version}: operational endpoints
\end{itemize}

\subsection{JSON-RPC handling and input extraction}
The JSON-RPC handler accepts only \texttt{message/send}. A practical compatibility point is that UI requests encode user text in \texttt{params.message.parts[].text}, not \texttt{params.text}. The Hub extracts text in the following order: (i) any \texttt{kind="text"} element in \texttt{params.message.parts}, falling back to (ii) \texttt{params.text} if necessary. This prevents input loss due to specification drift or client variance.

\subsection{UI compatibility mode and exception containment}
To prevent fatal UI failures, JSON-RPC responses always return a single text part. Structured data and citations are excluded from JSON-RPC responses. Additionally, the Hub encloses routing and downstream invocation in a protective \texttt{try/except} layer (a \texttt{safe\_route\_query} wrapper), ensuring that exceptions are rendered into a text summary instead of HTTP 500.

During development, we observed Cloud Run failures caused by application logic errors, such as a startup \texttt{NameError} and a \texttt{ValueError: too many values to unpack} due to a changed return signature of the routing function. These errors previously surfaced as UI “answer failed” messages; the containment layer prevented such failures from breaking the UI path.

\subsection{REST tool API for structured inspection}
The \texttt{POST /tools/query} endpoint supports UI-independent testing. It returns routing decisions, the downstream agent used, structured response payloads, and citation metadata as JSON. In evaluation, we queried the Hub via both JSON-RPC and REST to confirm that UI-compatible text responses and structured debug signals were cleanly separated.

\subsection{Downstream invocation and boundary-aware authentication}
All downstream agents are invoked via HTTP POST using A2A JSON-RPC. For public cross-project agents, requests are sent without an \texttt{Authorization} header (Cloud Run \texttt{--allow-unauthenticated}). For cross-account agents protected by IAM, the Hub fetches an OIDC ID token using its service account and attaches \texttt{Authorization: Bearer <token>}. We confirmed that an incorrect audience yields HTTP 401, while missing Invoker permissions yields HTTP 403.

\subsection{RAG route and observed permission failures}
For the DocQA route, the Hub first retrieves results via Discovery Engine. When needed for precise extraction (deadlines, numeric constraints), the Hub fetches the original document text from GCS. Initially, the Cloud Run execution service account lacked \texttt{storage.objects.get}, producing HTTP 403 during GCS reads and causing incomplete answers despite successful search. Granting the permission restored evidence-backed extraction and stable DocQA behavior.

\section{Evaluation}
We evaluate whether the proposed Hub operates stably under practical Gemini Enterprise UI conditions, focusing on: (i) correctness of deterministic routing, (ii) UI compatibility, and (iii) cross-boundary connectivity.

\subsection{Benchmark design}
We use four representative queries:
\begin{enumerate}[leftmargin=1.8em]
  \item \emph{Expense}: ``What is the expense reimbursement submission deadline?''
  \item \emph{PM support}: ``List three tasks for creating a project WBS.''
  \item \emph{General knowledge}: ``What is the height of Mount Fuji?''
  \item \emph{DocQA}: ``What is the deadline for notifying the infrastructure team for a P-1 incident?''
\end{enumerate}
They cover business policy Q\&A, PM assistance, public facts, and deadline extraction from internal documents. We evaluate both (a) JSON-RPC text-only responses (UI-compatible path) and (b) structured REST tool responses.

\subsection{Routing and response validity}
For all four queries, the Hub routed to the intended path: expense queries to the public cross-project agent, PM queries to the cross-account IAM-protected agent, general knowledge to the Hub’s Vertex AI route, and deadline queries to DocQA. On the JSON-RPC path, every response was a single text part and produced no UI rendering errors, indicating that text-only compatibility is effective under \texttt{acceptedOutputModes=[]}. On the REST path, routing outcomes, downstream agent identifiers, structured results, and citations were available, demonstrating separation between UI stability and internal observability.

\subsection{Evidence-backed extraction on the RAG path}
For the deadline query, DocQA returned the deadline ``within 15 minutes of incident detection'' along with a citation reference to the source document. We observed that without \texttt{storage.objects.get} on the Cloud Run service account, the system could retrieve search results but could not reliably extract the explicit deadline from the source text. After granting the permission, the same query produced stable evidence-backed extraction.

\subsection{Boundary-wise connectivity}
Within the same project, Vertex AI and Discovery Engine were usable through ADC without explicit token code. For a public cross-project agent, unauthenticated HTTPS invocation succeeded consistently. For the cross-account agent, invocation succeeded only when the Hub used an OIDC ID token with a correct audience and when Cloud Run Invoker permission was granted; misconfiguration produced 401/403 as expected. These results confirm that authentication is a discontinuous function of boundary type, even when the application-level A2A calls remain the same.

\section{Discussion / Lessons Learned}
Our study suggests that agent interoperability is dominated less by protocol specifications than by client UI behavior and cloud boundary conditions.
The experiments were conducted using Japanese natural-language queries in the
actual enterprise UI; however, query examples are presented in English for
readability.

\paragraph{Lesson 1: Prioritize UI compatibility.}
Real Gemini Enterprise UI requests include \texttt{acceptedOutputModes=[]} and provide input via \texttt{params.message.parts[].text}. Returning structured responses on the JSON-RPC channel can trigger UI errors. Enforcing text-only responses on the UI path while separating structured debug outputs into a REST API is an effective pattern for maintaining UI stability without losing testability.

\paragraph{Lesson 2: Authentication is the hidden determinant.}
Because A2A does not specify authentication, deployments inherit the complexity of Cloud Run and IAM boundaries. In our experiments, same-project calls were nearly transparent via ADC, cross-project public agents worked over unauthenticated HTTPS (with obvious security trade-offs), and cross-account invocations required explicit OIDC ID tokens and Invoker grants. This discontinuity is easy to overlook and is not emphasized in typical high-level documentation.

\paragraph{Lesson 3: Hub-based orchestration contains failures.}
A Hub can normalize and contain downstream failures (agent exceptions, IAM misconfigurations, partial RAG failures) into UI-compatible responses, preventing end-user experience collapse. In particular, shielding Cloud Run startup or application-logic errors from surfacing as UI-level failures is crucial for operational stability.

\paragraph{Lesson 4: Retrieval and evidence access are distinct.}
Even when managed search succeeds, evidence-backed extraction can fail if direct access to source documents (e.g., in GCS) is blocked by IAM. RAG deployments should explicitly treat evidence access as a first-class requirement, not an implicit consequence of retrieval.

\section{Conclusion}
We implemented and validated a Cloud Run--based A2A Hub that enables Gemini Enterprise UI to orchestrate agents and tool paths across same-project, cross-project, and cross-account boundaries. Our results clarify practical requirements that are difficult to infer from specifications alone, including strict text-only UI compatibility, boundary-dependent authentication design, and IAM considerations for evidence-backed RAG.

Evaluation on a four-query benchmark confirmed deterministic routing and stable UI responses via a text-only compatibility mode, while a separate REST tool API preserved structured inspection. We further showed that the Hub can absorb boundary differences and that cross-account connectivity depends on explicit OIDC ID tokens and Cloud Run Invoker permissions. These findings provide actionable guidance for future cross-platform and cross-organization agent integrations.

\bibliographystyle{unsrt}
\bibliography{references}

\end{document}